\documentclass{kluwer}
\usepackage{graphicx}
\begin{document}
\begin{article}
\begin{opening}
\title{\bf THE LIGHT CURVE OF V605 Aql -- THE \\
``OLDER TWIN'' OF SAKURAI'S OBJECT}

\author{HILMAR W. \surname{DUERBECK}\email{hduerbec@vub.ac.be}}
\institute{WE/OBSS, Brussels Free University (VUB), Pleinlaan 2, B-1050 
Brussels, Belgium}
\author{MARTHA L. \surname{HAZEN}\email{mhazen@cfa.harvard.edu}}
\institute{Harvard College Observatory, Cambridge, MA 21210, USA}
\author{ANTHONY A. \surname{MISCH}\email{tony@ucolick.org}}
\institute{Lick Observatory, University of California, Santa Cruz, CA 94065, 
USA}
\author{WALTRAUT C. \surname{SEITTER}\email{seitter@uni-muenster.de}} 
\institute{Muenster University, Germany}                               

\runningtitle{THE LIGHT CURVE OF V605 Aql}
\runningauthor{DUERBECK et al.}

\begin{ao}
Hilmar W. Duerbeck\\
WE/OBSS\\
Vrije Universiteit Brussel\\
Pleinlaan 2\\
B-1050 Brussel\\
Belgium
\end{ao}

\begin{abstract} 
``Nova Aquilae No.\,4'', discovered in 1919, later renamed V605 Aql, 
was recognized to be a final helium flash object only in the 1980s. 
Clayton and De Marco (1997) gave a detailed description of the available 
spectroscopic and photometric material. Here we try to re-analyze 
the photometric record. The photographic material is still 
available and has been used to construct a revised light curve. Only fragments
of the visual observations were published; the remainder 
appears to be lost.
\end{abstract}

\keywords{Stars: individual: V605 Aql (N Aql No. 4 (1919)); 
V4334 Sgr (Sakurai's object) -- post-AGB evolution -- final He flash objects
-- photometry}

\end{opening}

\section{Introduction}
Nova Aquilae No.\,4 was discovered by Max Wolf 
on photographic plates of the
Heidelberg Observatory taken in 1919, and was announced in 
a short note in the {\it Astronomische Nachrichten} of 1920 May 28 (Wolf
1920). Subsequently, the star was analyzed on Harvard plates, and another 
short note was published by Ida Woods (1921). At that time,
the object was thought to be a very slow nova.
Its peculiarity was only noted when Lundmark (1921) 
took spectra in September 1921,
and classified it as a carbon star. This classification caused little 
interest at the time of publication, and lead to no closer studies. 
Visual observations, carried out at later stages, showed pronounced fadings 
as well as recoveries, and the star was thought to be a long period variable.

\section{Observations}
The observational record can be classified as follows. 
Photographic
observations in Heidelberg (magnitudes and comparison stars),
part of the photographic observations in Harvard, and all
photographic observations at Lick were published. All estimates and comparison
stars at Harvard could be recovered in Miss Woods' notebooks.
They contain, in addition to the published record, many negative
observations (i.e. times when the object was not visible above
a certain magnitude limit), and estimates made after the 1921 paper. 

All photographic observations were transformed
to the modern magnitude scale, using data from the Tycho-2 catalogue. 
The Lick observations were newly estimated, using the Harvard
scale (they agree very well with Lundmark's old estimates), 
and transformed into Tycho-2 magnitudes. The differences
between the original (pg) and the modern ($B_T$) values were $1^{\rm}$ 
and $0.5^{\rm m}$, for Heidelberg and Harvard, respectively.

The visual magnitudes are less known and more difficult to use. 
The two observers,
Michael Esch SJ (Sternwarte Valkenburg) and Kasimir Graff 
(Hamburger Sternwarte), 
published only short accounts in the
Beobachtungs-Zirkular of the Astronomische Nachrichten, 
without giving information on standard stars
(Esch 1923a,b; Graff 1923a,b).
Both authors had planned more detailed publications.
During the years  1920 -- 1932,
Esch collected 49 observations 
(Esch 1935), but it is possible that the major part of them were negative
ones.
In the 1930s, Esch began to publish his extensive observations of variable
stars in the Ver\"offentlichungen der Sternwarte des Ignatiuskollegs Valkenburg
(The Netherlands);
this series was interrupted by his death (1938). In 1942, 
actions against the Jesuits took place. The 
Kolleg was devastated and dissolved by German secret police.  
Fortunately, after Esch's death, his notebooks were taken to the 
Vatican Observatory for
evaluation. A first (and only) publication by Albert Zirwes, 
appeared in 1943 in the Publications
of the Vatican Observatory. V605 Aql was, however, not among the published
stars, and a recent search for Esch's observing book(s) in the archives of the 
Vatican observatory has been unsuccessful so far.

Graff also planned to publish visual magnitude scales of variables and
associated observations that he had carried out in Hamburg (Graff 1928). 
Then, he accepted a call 
as director of the Vienna Observatory, and his project was shelved.
He had to resign at the time of the Nazis, but was re-instated after 
the end of WW II.
By this time, Graff was a sick man and died soon after. His written estate 
appears to be lost.

\section{Results}
The fragmentary light curve of V605 Aql (Fig. 1) is tantalizing. 
The photometric calibration cannot be regarded as final, it supersedes, 
however, the published light curve of Harrison (1996).

\begin{figure}
\begin{center}
\includegraphics[width=100mm,angle=270]{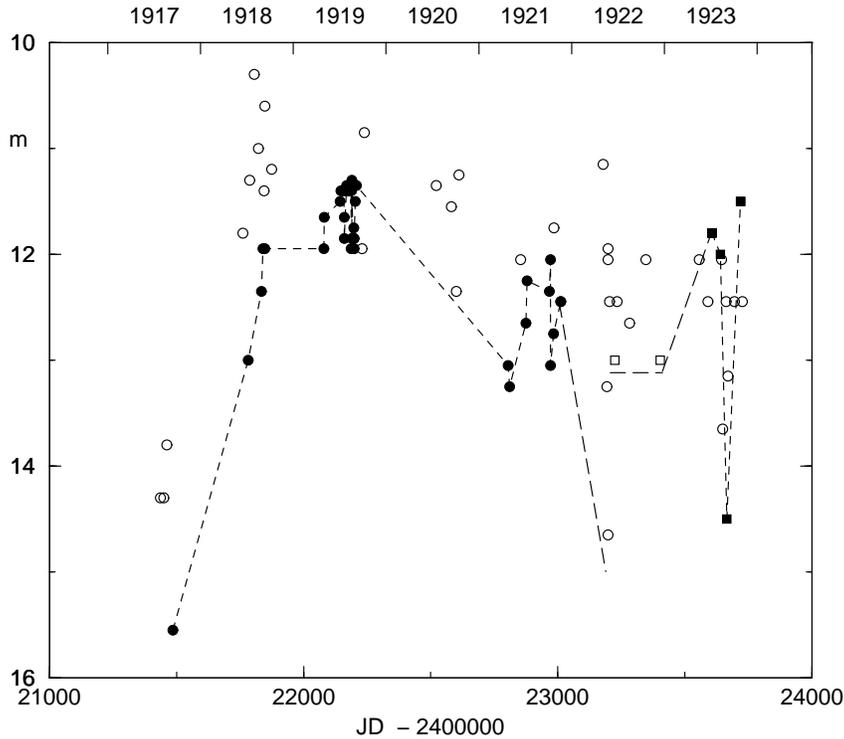}
\caption[]{The light curve of V605 Aql 1917 -- 1923. 
Filled circles: photographic
magnitudes, open circles: upper limits of photographic magnitudes, filled
squares: visual magnitudes, open squares: upper limits of visual magnitudes.
The photographic and the visual light curves are connected by dashed lines.
Long-dashed lined indicate uncertain light curve segments, which use
upper limits of negative observations}
\end{center}
\end{figure}

The rise to photographic maximum and the early decline of V605 Aql are 
fairly well documented. In view of
the many negative photographic observations at late phases
(also indicated by an upper limit in early 1923), the 
positive visual observations indicate that the object was very red 
at this time. 
No `quasi-simultaneous' photographic and visual 
observations exist. Since
rapid, large light fluctuations are documented, no colour index
can be derived with certainty.

The photographic data show that V605 Aql rose
to maximum  with a rate of about $-0.01$ mag/day, and 
afterwards declined with $+0.001$ mag/day. These values 
are uncertain, but are similar to those observed in 
V4334 Sgr ($-0.015$ mag/day and $+0.001$ mag/day). 

After 1923, V605 Aql disappeared from view:  no positive observations down to
$16 - 17.5$ mag exist for the years 1928 -- 1979 (Fuhrmann 1981). 
 
A comparison of the `period of visibility' of both objects
indicates that V4334 Sgr vanished more quickly from view. 
Unfortunately, the total
time of visibility of V605 Aql cannot be stated with certainty, unless 
Esch's notebook is recovered in the Vatican archives. In any case, if the
analogy between V605 Aql and V4334 Sgr holds, Sakurai's object will remain
very faint for some decades.

\section{Conclusions}
V4334 Sgr and V605 Aql appear to have very
similar light curve characteristics, 
and V605 Aql might serve as a `pathfinder' for the
events that will happen to V4334 Sgr in the future, e.g. the gradual
re-appearance of the hot central star within the dust cocoon.

\section*{Acknowledgements}
We thank Dr. A. Schnell (Vienna Observatory) and Dr. P. Maffeo SJ (Vatican
Observatory) for information on Graff's and
Esch's estates, and A. van Genderen (Sterrewacht Leiden), Br. M. Pillat SJ 
(Archiv der Norddeutschen Provinz der Jesuiten K\"oln), and the 
Bundesarchiv (Koblenz) for tracing the fate of the Ignatiuskolleg.

\end{article}
\end{document}